\documentclass{article}
\usepackage{spconf}
\usepackage{amsmath,graphicx}
\usepackage{tipa}
\usepackage{xspace}

\usepackage[font=small,skip=9pt]{caption}

\usepackage{hyperref}
\hypersetup{
    pdftoolbar=true,
    pdfmenubar=true,
    pdffitwindow=false,
    pdfstartview={FitH},
    pdfpagemode=FullScreen,
    pdftitle={Crowdsourced multilingual speech intelligibility testing},
    pdfauthor={Laura Lechler, Kamil Wojcicki},
    pdfsubject={Crowdsourced multilingual speech intelligibility testing},
    pdfcreator={Laura Lechler, Kamil Wojcicki},
    pdfproducer={Laura Lechler, Kamil Wojcicki},
    pdfkeywords={multilingual} {crowdsourcing} {speech intelligibility assessment} {diagnostic rhyme test} {DRT},
    pdfnewwindow=true,
    colorlinks=true,
    linkcolor=blue,
    citecolor=blue,
    filecolor=blue,
    urlcolor=blue,
    breaklinks=true
}
\graphicspath{ {./figures/} }

\def\eg{{{e.g.,}\xspace}}
\def\ie{{{i.e.,}\xspace}}
\def\ttest{{\textit{t}-test}\xspace}
\def\ci95{{CI$_{95}$}\xspace}

\usepackage[noadjust]{cite}

\usepackage[pscoord]{eso-pic}

\usepackage{background}
\SetBgAngle{0}
\SetBgScale{1}
\SetBgColor{black}
\SetBgOpacity{1}
\SetBgPosition{.51\paperwidth,-1.08\textheight}
\SetBgContents{
{\parbox{\paperwidth}{%
\footnotesize \copyright 2023 IEEE.  Personal use of this material is permitted. 
Permission from IEEE must be obtained for all other uses, in any current or future media,\\
including reprinting/republishing this material for advertising or 
promotional purposes, creating new collective works, for resale or redistribution\\
to servers or lists, or reuse of any copyrighted component of this work in other works. 
DOI: \href{https://ieeexplore.ieee.org/document/10447869}{\textit{10.1109/ICASSP48485.2024.10447869}}.\\
\mbox{}\\
     \mbox{}\\}
     }%
}

\makeatletter
\newcommand\footnoteref[1]{\protected@xdef\@thefnmark{\ref{#1}}\@footnotemark}
\makeatother

\title{Crowdsourced multilingual speech intelligibility testing}
\name{Laura Lechler, Kamil Wojcicki}
\address{Cisco Systems, Inc.} 

\begin{document}

\ninept
\maketitle
\begin{abstract}

With the advent of generative audio features, 
there is an increasing need for rapid evaluation of their impact 
on speech intelligibility.
Beyond the existing laboratory measures, 
which are expensive and do not scale well,
there has been comparatively little work on 
crowdsourced assessment of intelligibility. 
Standards and recommendations are yet to be defined,
and publicly available multilingual test materials are lacking.
In response to this challenge, 
we propose an approach for a crowdsourced intelligibility assessment.
We detail the test design,
the collection and public release of the multilingual speech data, 
and the results of our early experiments.
\end{abstract}
%

\section{Introduction}
\label{sec:introduction}

Methods for the assessment of speech quality and intelligibility
are indispensable during both algorithm development
and product release testing.
They should enable the algorithm researchers to iterate at speed
in a guided and efficient manner. 
As such, they should be accurate and reliable 
but also scalable, \ie~rapidly deployable at scale and at a permissible cost.
Meeting these potentially conflicting requirements poses a challenge.

Objective measures are often used, 
especially in the early stages of algorithm development,
as they can be applied readily and cheaply.
In the past, 
objective measures that are intrusive in nature, 
i.e., those that require a reference 
(\eg~\cite{torcoli_objective_2021, taal_algorithm_2011}), 
have proven especially useful in practice. %
In recent years, however, audio algorithms based on
generative approaches have been gaining popularity
(\eg~\cite{valin_real_time_2022, richter2023speech, lemercier_storm_2023}).
These algorithms aim to predict a plausible audio instance,
rather than a specific reference.
For such methods, 
non-intrusive measures are needed instead, 
i.e., those that can blindly assess a speech signal
without requiring an external reference 
(\eg~\cite{torcoli_objective_2021, feng_nonintrusive_2022}).
Non-intrusive assessment, however, 
remains a challenging problem to solve effectively 
\cite{feng_nonintrusive_2022}.
Importantly, both intrusive and non-intrusive methods must be utilized with care,
as their predictions can be inaccurate (or invalid) under conditions differing 
from those assumed in their development and validation \cite{torcoli_objective_2021}.

Listening tests remain the gold standard 
for assessing quality and intelligibility of speech \cite{Loizou.13.Speech}.
Traditionally, 
such tests have been conducted in laboratories 
under carefully controlled and standardized conditions. 
However, this is both costly and time consuming 
and does not lend itself to rapid and scalable testing.
This, in turn, inhibits the speed at which algorithm research can be carried out.

In recent years, 
crowdsourced assessments have gained popularity \cite{naderi_towards_2020}. 
Unlike laboratory tests, 
crowdsourced approaches scale very well,
given their relatively low cost and the immediate access
to a large pool of na\"ive listeners across the world.
In these methods, 
the reduced control over experimental conditions 
could lead to increased bias and variance.
Averaging over a large number of responses
and a specialized test design are used to ensure accuracy and reliability.

Crowdsourced assessments of speech quality 
have been generally utilized with success (\eg~\cite{dubey2023icassp}).
They have been well covered in the 
literature (\eg~\cite{naderi_open_2020, naderi_speech_2021_article, suarez_crowdsourcing_design_2023}), 
standardized \cite{itu_p808_2021}, 
open-sourced in terms of software 
and with public test data releases (\eg~\cite{naderi_open_2020}). 
The above is in a stark contrast to
the crowdsourced assessment of speech intelligibility.
While there are recent publications on this topic
(\eg~\cite{voran_crowdsourced_2017, chua_quantifying_2017, yamamoto_comparison_2021, wolters_crowdsourcing_2017, yamamoto_effective_2022}), 
the literature is generally scarce, and
standards and recommendations are yet to be proposed.
Text materials (\eg~word or sentence lists) across a multitude of languages
are not efficiently summarized in a single reference
and in some cases difficult to access.
Notably, 
publicly available multilingual audio recordings for testing are not generally available.
The conditions described above create a barrier for the effective assessment of speech intelligibility.

Current speech algorithm research is moving towards not previously achievable benefit heights, 
for example, through generative reconstruction of missing speech content 
(\eg~\cite{valin_real_time_2022, richter2023speech, lemercier_storm_2023}).
We may soon see single-channel speech enhancement algorithms 
that markedly improve intelligibility
in challenging real-world environments. 
With these new potential benefits come also additional risks. 
For example, 
while the generated speech content could lead to 
intelligibility improvements in some phoneme categories,
it may lead to degradations in others (\eg~/f/ reduced to /p/). 
These degradations might still resemble real speech, 
but could be inappropriate in a given context, 
which non-intrusive measures may not detect.

Furthermore, the improvements and degradations could vary across languages.
The researchers must have accessible and effective tools 
to assess the effect of these algorithms on speech intelligibility.

In the dawn of powerful data-driven 
(yet largely black-box) audio algorithms, 
the need for accurate, reliable, and scalable multilingual assessment
of \textsl{both} quality and intelligibility is thus pressing.
The present work aims to be a step forward in addressing this need.
Specifically, 
we detail a crowdsourced test for intelligibility assessment,
contribute multilingual data as a public release for the research community, 
and present results of early experiments.

The remainder of this work is organized as follows.
A brief overview of some of the existing 
approaches to speech intelligibility assessment 
is presented in Section~\ref{sec:si-assessment}.
The proposed test design is detailed in Section~\ref{sec:test-design}.
Experiments and results are given in Section~\ref{sec:experiments-and-results}.
The discussion and future work are presented in Section~\ref{sec:discussion}.
Conclusions are stated in Section~\ref{sec:conclusion}.

\section{Speech intelligibility assessment}
\label{sec:si-assessment}

A variety of laboratory tests have been proposed for speech intelligibility assessment, 
including tests based on nonsense syllables, words, and sentences \cite{Loizou.13.Speech}.
Standardized language-specific tests have been established \cite{acoustical_society_of_america_method_2020}, 
such as those based on phonetically balanced word lists used in transcription tasks (\eg~measuring word error rate)
and those focused on closed-set identification tasks. 
Examples for closed-set tasks are the diagnostic rhyme test (DRT) \cite{voiers_evaluation_1965} and 
the modified rhyme test (MRT) \cite{house_articulationtesting_1965}. 
While transcription tasks, 
including matrix tests with semantically unpredictable sentences (\eg~\cite{kollmeier_multilingual_2015}), 
offer the advantage of being able to test entire sentences, 
word-based multiple-choice recognition tasks are easier to conduct, 
as spelling errors and other text input challenges are avoided. 
Sentence transcription tasks are also thought 
to be influenced by considerable memory effects, 
preventing the re-use of materials with the same participant group. 
This is particularly hard to control in crowdsourcing scenarios, 
where different research groups may access the same participant pool. 
Therefore, new controlled materials are required frequently. 
Although approaches such as keyword spotting aim to facilitate this
\cite{valentini-botinhao_efficient_2023}, it remains a non-trivial task. 

We note that previous efforts to assess speech intelligibility 
by means of crowdsourcing often considered transcription tasks 
(\eg~\cite{chua_quantifying_2017, wolters_crowdsourcing_2017, yamamoto_comparison_2021, yamamoto_effective_2022}).
A subset of these focused on text-to-speech (TTS) algorithm evaluation. 
As one major challenge of TTS systems is the appropriate modelling of prosody, 
which may be best evaluated over longer periods, this focus is understandable. 
However, 
particularly (though not exclusively) for speech codec and speech enhancement algorithms, 
intelligibility can be impaired at the phonemic level of granularity. 
As rhyme tests do not include semantic clues to the correct solution, 
they may be more sensitive to short-term (phoneme-level) degradations. 
When aiming for standardized, repeatable tests in a crowdsourcing environment, 
the reuse of testing materials is advantageous. 
To the best of our knowledge, 
the only publicly available word-based audio dataset 
for a closed-set testing is the MRT in English \cite{voran_using_2013}, 
despite the availability of word lists in many languages for both the MRT and DRT approaches.%
\footnote{\label{repo}%
We make available the recorded speech materials for the five languages reported here, 
as well as an extended bibliography pointing to previously published DRT word lists in 16 languages, 
via the following repository:\newline 
\url{https://github.com/cisco/multilingual-speech-testing}}

Both DRT and MRT have been shown to have good correlation with each other 
and with other intelligibility tests \cite{kryter_comparisons_1965}, 
and have been shown to yield repeatable, 
self-sufficient results \cite{williams_selecting_1967,williams_relation_1968}. 
However, 
as the DRT presents only single words in isolation
along with two transcription options instead of six, 
the testing time of the DRT is much shorter. 
More rapid testing allows researchers to collect more responses 
and include instances of the test items from more speakers, 
while keeping costs manageable. 
This is particularly relevant for a crowdsourcing setting.
Further, 
the linguistic and acoustic insight of the DRT, 
with test items belonging to classes of distinctive linguistic features 
which are acoustically interpretable, 
poses a useful tool for both development and benchmarking. 
For these reasons and due to the ample availability of DRT text materials in a variety of languages, 
we decided to employ the DRT for the crowdsourced intelligibility assessment 
detailed in the remainder of this work.

\section{Crowdsourcing test design}
\label{sec:test-design}

The DRT was implemented as a repeatable online survey using Qualtrics 
and deployed on crowdsourcing platforms such as Prolific. 
Generally, 
care was taken to adhere to standards and recommendations as closely as possible 
\cite{itu_p807_2016, acoustical_society_of_america_method_2020}. 
However, 
the adaptation to the crowdsourcing environment implied limitations on 
the direct control over factors such as 
the listening environment and equipment used, 
as well as focus and hearing capabilities of the participants. 
This was mitigated 
with pre-screening techniques and post-filtering of the results through validation mechanisms. 
These approaches are based on prior work 
on now widely practiced crowdsourced speech quality evaluations 
\cite{itu_p808_2021, naderi_open_2020}.

\subsection{Word lists}
\label{sec:word-lists}

We gathered previously published word lists for the DRT in the following languages:
English \cite{itu_p807_2016}, 
Spanish \cite{de_cardenas_cuaderno_1994},
French \cite{peckels_test_1973}, and 
Mandarin Chinese
\cite{mcloughlin_subjective_2008}.
We also created a comparable word list for the German language \cite{lechler_german}. 
Word lists for further languages will be considered in future work.\footnoteref{repo}

For English, we chose the DRT word list recommended by the ITU \cite{itu_p807_2016}, 
which consists of both DRT and diagnostic alliteration test items, 
contrasting consonants at either the beginning or the end of a word.
In the following, 
we use the term DRT in a more general way irrespective of the location of the contrast.
For Spanish \cite{de_cardenas_cuaderno_1994}, the first set of sibilation items were excluded, 
as they contain the /s/ vs. /\textipa{T}/ contrast that only exists in European Spanish.
Separate word lists for consonant and tonal contrasts were included for Chinese \cite{mcloughlin_subjective_2008}.

\subsection{Test sets}
\label{sec:test-sets}

Audio stimuli for the test sets corresponding to the word lists outlined in Section \ref{sec:word-lists} 
were collected via crowdsourcing from native speakers of each language.

The initial recordings were curated to ensure sufficient quality:
mispronounced, unintelligible, noisy, and low-quality 
recordings were excluded from further consideration.
The curated files were downsampled to 16 kHz sampling rate, 
cropped to include 500 ms leading and trailing silence,
faded in and out at the onsets and offsets, respectively, to avoid discontinuities,
and then level-normalized to --26 dB RMS.
The final gender-balanced test set for each language included 
six instances of each target word (three female, three male),
with different sets of talkers across words.

\subsection{Test blocks}
\label{sec:test-blocks}

The word list recordings were split into 6 or 12 pseudo-randomized blocks, depending on the number of word pairs on the list. 
Each block contained one word of each word pair in the test list (\eg~96 words/pairs for English).
Each block was balanced for speaker gender and the presence or absence of the distinctive features. 
The order of items within a block was randomized at test time. 
Each test session evaluated one single treatment condition.
The 16 kHz sampled clean wideband recordings 
for a given language, 
are henceforth referred to as the clean wideband, or WB, condition.

\subsection{Motivating rewards}
\label{sec:rewards}
Participants received rewards corresponding to an \textsl{hourly} rate exceeding \$8.00 USD on average. 
To encourage good performance and attention, 
top participants were rewarded with a bonus of \$0.10 USD per \textsl{session}.
The performance was assessed based on the number of validation questions answered correctly.

\subsection{Pre-screening and participants}
\label{sec:pre-screening}

Access to the study on Prolific %
was restricted to \textsl{a priori} self-reported first-language speakers of the target language, 
who stated to reside in a country where this language is commonly spoken, 
did not have diagnosed or suspected dyslexia nor hearing problems, 
and had an approval rate of $\geq$98\%. 
The use of headphones was specified as mandatory.
A gender-balanced sample of participants was selected. 
Panels of 20 listeners on average were recruited. 
The smaller Spanish and Chinese tonal DRTs required only six panels, while other tests required twelve panels.

\subsection{Test Procedure}
\label{sec:response-quality}
 
The participant pool was selected based on pre-screening criteria described in Section~\ref{sec:pre-screening}.
Informed consent was obtained from eligible participants 
who were presented with the experiment description. 
This information specified the requirement to listen to 
recordings over headphones in a quiet environment.
Consenting participants then proceeded to a questionnaire, 
which repeated the screening questions to ensure consistency
between current self-reported information
and responses previously recorded by Prolific.
Headphone use was also validated through the questionnaire. 
Additionally, 
participants were asked to state their age group and self-identified gender, 
if they wished to do so.

The questionnaire was followed by a speech perception test in the form of digits in noise.
Six randomized digit triplets at varying signal-to-noise ratios were presented.
The English samples were taken from the P.808 toolkit \cite{naderi_open_2020},
while other language versions were created in a comparable fashion. 
All speaker voices were male. 
The threshold was set to identify participants with normal hearing.

Participants were presented with a practice run consisting of 16 examples 
from the test set in the WB condition. 
They were asked to select which word they heard from 
a two word alternative choice displayed on their computer screen. %
The training session served multiple purposes: 
ensuring any required audio output level changes on the playback device could be made ahead of testing; 
familiarizing participants with the test format and procedure; 
confirming participants understood the task and could correctly 
identify the majority of words from the baseline condition.
Additionally, 
20 catch trials were included in each testing block.
WB recordings were utilized as validation questions to 
gauge participant attention and performance.
Only submissions that fulfilled the following criteria were included in the data analysis: 
participants had fulfilled the pre-screening criteria, 
reported to be using headphones in a quiet environment, 
passed the digit perception in noise test, and 
had solved more than 80\% of the validation questions correctly.

\begin{figure} [!t]
    \centering
    \includegraphics[width=0.435\textwidth]{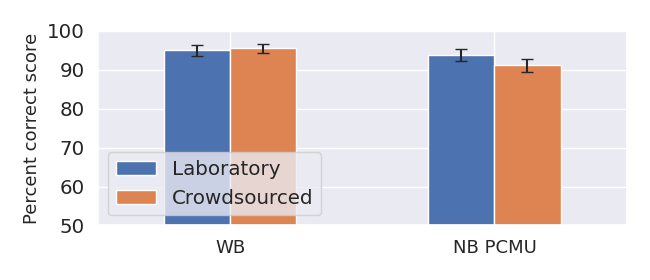}
    \caption{Laboratory and crowdsourced intelligibility scores for Spanish WB and NB PCMU conditions.}
    \label{fig_ex1}
\end{figure}

\subsection{Score calculation}
\label{sec:score}
Intelligibility scores were calculated following 
the formula reported in \cite{itu_p807_2016}: $P(c)=[R-W]/(R+W)*100$, 
where $P(c)$ is the percent (\textit{P}) correct (\textit{c}) score (adjusted for guessing), 
$R$ the number of correct responses, and 
$W$ is the number of incorrect responses. 

Due to the larger number of judgements per file (approximately 20 responses on average) 
compared to traditional laboratory tests 
(typically 8--9 \cite{acoustical_society_of_america_method_2020,itu_p807_2016}), 
we calculated the scores for each audio sample 
rather than across participant trials. 
This allows us to assess score changes on a per-file, target word, or phoneme basis 
in addition to averaging across the whole test set and distinctive features. 
We report the mean and 95\% confidence interval (CI$_{95}$).

\section{Experiments and results}
\label{sec:experiments-and-results}

\subsection{Experiment 1: Crowdsourced accuracy on Spanish}
\label{sec:experiment-1}

To assess the accuracy of the crowdsourced test, 
we conducted a controlled in-house experiment involving expert listeners. 
We conducted this test on the Spanish WB 
and narrowband with the PCMU (G.711 $\mu$-law) codec applied (NB PCMU).
All listeners (9 female, 7 male) were native speakers of Spanish. 
The test interface was the same as in the crowdsourced tests. 
The controlled listening experiment was performed in a quiet environment 
and using high-quality wired headphones. 
Internal participants evaluated three randomly allocated blocks of test files 
(half the test set) for each of the two conditions. 
The two conditions were evaluated in separate tests and in separate testing sessions 
with sufficient time between the two tests. 
Each internal session took approximately 45 minutes. 
Crowdworkers evaluated one block per study. 
The median completion time for these sessions was 17 minutes. 
Statistical differences between mean scores obtained for different conditions 
were assessed using a \ttest at a significance level of $p$$<$0.05, 
unless reported otherwise. 

We compared these internal results to the crowdsourced results 
obtained on Prolific (Fig.~\ref{fig_ex1}). 
In the internal group, 
the drop in overall intelligibility by 1.2 score points in the NB PCMU condition compared to WB was not significant. 
In the crowdsourcing group, 
the NB PCMU score was 4.3 points below the WB reference score. 
This difference was statistically significant. 
There was no statistically significant difference between 
the two participant groups in evaluating the WB condition. 
However, crowdworkers scored 2.6 points lower 
than the expert listeners in the NB PCMU condition, 
which was a statistically significant difference.

\subsection{Experiment 2: Crowdsourced consistency on Spanish}
\label{sec:experiment-2}

To assess the consistency of the crowdsourced test 
we repeated the first trial assessing NB PCMU in Spanish: 
(a) with the same participant group and 
(b) with a different participant group. 
Out of the 150 participants from the first trial, 
127 participated again in the second trial. 
A new set of 150 participants was recruited for the third trial.

\begin{figure}[t!]
    \centering
    \includegraphics[width=0.435\textwidth]{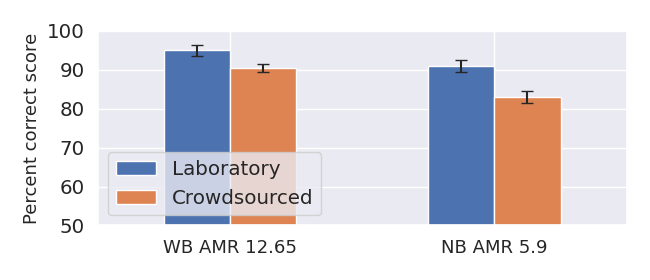}
    \caption{Overall intelligibility scores for English obtained for two codecs under laboratory conditions \cite{itu_p807_2016} and via crowdsourcing.}
    \label{fig_ex3a}
\end{figure}

In the first trial, 
an overall intelligibility score of 91.2 (±1.7 {\ci95}) was obtained. 
In the second trial, 
an overall intelligibility score of 92.4 (±1.5 {\ci95}) was achieved. 
The difference between the two trials was not statistically significant in a \ttest. 
A good test-retest consistency was confirmed by Pearson correlation ($\rho$$=$0.87, $p$$<$0.001). 
The third participant group achieved an overall intelligibility score of 91.4 (±1.6 {\ci95}). 
The difference between the first and the third trial was also not statistically significant. 
Trials 1 and 3 were also well correlated ($\rho$$=$0.86, $p$$<$0.001).

\subsection{Experiment 3: Codec comparisons on English}
\label{sec:experiment-3}

\begin{figure}[!t]
    \centering
    \includegraphics[width=0.435\textwidth]{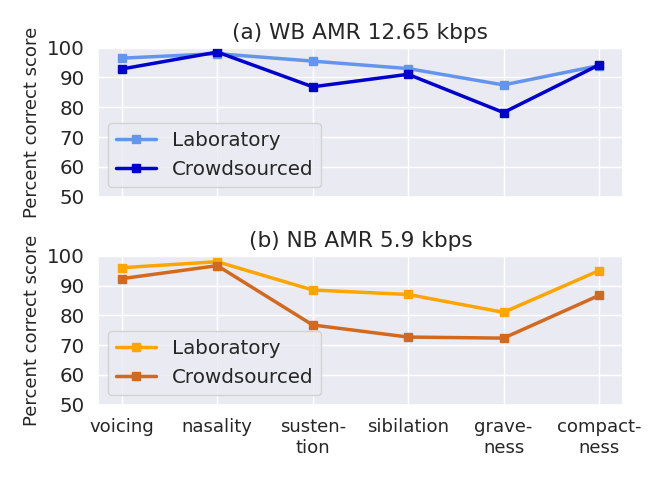}
    \caption{Intelligibility scores for English for two codecs under laboratory \cite{itu_p807_2016} and crowdsourcing conditions per distinctive feature.}
    \label{fig_ex3b}
\end{figure}

To compare our crowdsourced results to results reported in the literature, 
we evaluated two codecs with reference results for the same test set. 
The two codecs tested are AMR-WB at 12.65 kbps and AMR-NB at 5.9 kbps. 
The crowdsourced results were compared with laboratory results reported in the ITU recommendation in \cite{itu_p807_2016}.
In the overall score comparison (Fig.~\ref{fig_ex3a}), 
the expected pattern of degradation between the two AMR codecs 
is replicated by the crowdsourced tests. 
The crowdsourced absolute mean scores are lower than the laboratory scores, 
and the difference between the AMR WB codec and the NB codec is more pronounced. 
When comparing scores broken down by distinctive feature, 
similar trends between the categories can be observed (Fig.~\ref{fig_ex3b}). 
The mean scores for the distinctive features for the two test procedures are significantly correlated 
(WB: Pearson's $\rho$$=$0.86 ($p$$<$0.05); NB: Pearson's $\rho$$=$0.94 ($p$$<$0.05)).

\subsection{Experiment 4: Wideband vs. PCMU across languages}
\label{sec:experiment-4}

We conducted baseline experiments in 
English, German, Spanish, French, and Mandarin Chinese for two conditions: WB and NB PCMU.
The expected pattern of a intelligibility degradation caused by the PCMU codec 
can be observed across all languages for the consonant tests performed in the no-noise conditions. 
The relative change was statistically significant for all languages tested 
and came to approximately 4--5 mean score points. 
No statistically significant difference was observed for the Chinese tonal DRT section. 


\section{Discussion and future work}
\label{sec:discussion}

We conducted four experiments assessing the use of a crowdsourced DRT approach. 
Our results show that although intelligibility is more easily impaired 
in crowdsourced participants under degraded conditions, 
the relative scores are sensible, 
follow expected patterns, 
and show good correlations with laboratory studies. 
We observed this increased impact in Experiments 1 and 3, 
where crowdsourced results were compared with results obtained in controlled conditions. 
Potential factors that may have led to the greater observed degradations are, 
\eg~differences in headphone quality, environmental noise levels, 
attention levels, or hearing capabilities \cite{wolters_crowdsourcing_2017}.
The observation that relative scores obtained in crowdsourced tests 
are meaningful even though absolute scores may be lower than in laboratory experiments, 
is in line with other work on using crowdsourcing for speech perception and evaluation tasks 
(\eg~\cite{cooke_overview_2013, yamamoto_comparison_2021}).

\begin{figure}[!t]
    \centering
    \includegraphics[width=0.435\textwidth]{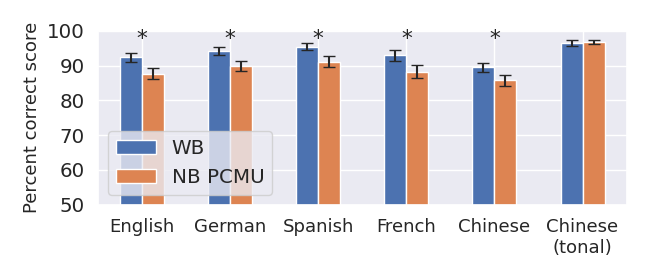}
    \vspace{-2mm}
    \caption{Overall intelligibility scores obtained for several languages. Asterisks indicate statistical significance at a level of $p$$<$0.05.}
    \label{fig_ex4}
\end{figure}

Further, 
we argue that the benefits of a more diverse, inclusive, and potentially more realistic participant pool 
outweigh the restricted means of experimental control. 
The good correlations with laboratory experiments and high intelligibility scores for the WB condition 
indicate a successful recruitment and filtering procedure.  
Our results obtained in Experiment 2 show good repeatability both within and between crowdsourced participants. 
This is in line with findings by Voran and Catellier \cite{voran_crowdsourced_2017}, 
who demonstrated that their crowdsourced implementation of the MRT 
had high agreement with and higher repeatability than laboratory tests. 
As we calculate overall averages from per-file percent correct scores, 
these correlations reflect the consistency of the test even at the smallest level of granularity.

We further demonstrated that the DRT consonant tests in the five languages tested here 
show similar rates of decrease in intelligibility for the PCMU codec. 
Further analysis shows that the majority of the highly confused word pairs rely 
on acoustic information contained in the higher frequency bands absent in the NB condition. 
However, as expected, 
the treatment did not cause any significant intelligibility impairments in the tonal section of the Chinese DRT, 
as tonal acoustic information is predominantly carried by pitch cues, 
which are preserved in the bandlimited signal \cite{zee_tone_1980}.


In this preliminary work, 
we did not report the effect of adding noise to shift the assessment 
into the steep part of the psychometric curve. 
Initial experiments involving the addition of speech-shaped noise (results not reported here) 
show the expected effect of lowering the intelligibility threshold and 
increasing the test's sensitivity to differences between treatments. 
We intend to demonstrate this in follow-up work. 
Initial test results (also not reported here) indicate considerable differences 
in the response quality across different crowdsourcing platforms, 
such as Prolific and Amazon Mechanical Turk, 
which we would like to investigate further.
Work currently underway will see the release of additional language data, 
\eg~Arabic and Japanese, 48 kHz versions of the test sets, and test software.

\section{Conclusion}
\label{sec:conclusion}

In this work, 
a repeatable and cost-efficient crowdsourced multilingual intelligibility assessment was proposed. 
The proposed assessment is an adaptation of a standardized speech intelligibility test 
to the crowdsourced context based on well-established principles from speech quality research. 
We detailed the test design, data collection and release, 
along with results of early experiments, 
demonstrating significant correlations to laboratory tests and good repeatability. 
It is hoped that the publicly available multilingual test sets 
will enable researchers to evaluate intelligibility performance in a multitude of languages 
and serve for benchmarking across studies.

\bibliographystyle{IEEEbib}
\bibliography{refs_final}

\end{document}